\documentstyle[aps,epsf]{revtex}

\def\nueff{N_{\nu}}

\begin{document}
\baselineskip 14pt

\title{Active-Sterile Neutrino Mixing in the
Early Universe and Primordial Nucleosynthesis}
\author{Kevork Abazajian, Xiangdong Shi and George M. Fuller}
\address{Department of Physics, University of California, San Diego, 
         La Jolla, California 92093-0350}
\maketitle

\begin{abstract}
We investigate the effects of matter-enhanced (MSW) transformation of
neutrinos in the Early Universe on the primordial helium yield
($Y_p$). We find that $Y_p$ is affected much more by the MSW-induced
alterations in the neutrino energy spectra than by the associated
change in expansion rate.  Specifically, the absence due to
transformation of low energy electron neutrinos can significantly
affect neutron-proton weak interconversion rates through the lifting
of Fermi-blocking of neutron decay at low energies and through halting
low-energy neutrino capture on neutrons.  We find that the change of
$Y_p$ within a causal horizon is $-0.005\le \delta Y_p \leq 0.013$ for
$m_{\nu_{\mu,\tau}}^2 - m_{\nu_s}^2 \le 10^4\, \rm eV^2$ in the case
of $\nu_{\mu,\tau}$-$\nu_s$-$\nu_e$ mixing, with the lower limit at
$m_{\nu_{\mu,\tau}}^2 - m_{\nu_s}^2 \approx 100\, \rm eV^2$, and
$-0.002\le \delta Y_p \leq 0.020$ for $m_{\nu_e}^2 - m_{\nu_s}^2 \le
1\, \rm eV^2$ in the case of $\nu_e$-$\nu_s$ mixing.

\end{abstract}

\section{Introduction}

Big Bang Nucleosynthesis (BBN) remains one of the most successful
probes of early times in the hot big bang cosmology.  Standard BBN
(SBBN) assumes the Standard Model proposition of only three massless
light neutrinos, leaving the primordial element abundances
characterized by only one parameter, the baryon-to-photon ratio
$\eta$.  Now, the evidence for neutrino mass from the Super-Kamiokande
atmospheric neutrino observations is overwhelming \cite{superk}.  In
addition, the solar neutrino deficit and the LSND experiment
\cite{solar} are hinting at the presence of a fourth light neutrino
that would necessarily be ``sterile'' due to the $Z$ decay width
\cite{zdecay}.  The role of neutrino masses and their mixing bring
another aspect to the physical evolution of the Early Universe, when
neutrinos played a large role.  The effects of neutrinos on big bang
nucleosynthesis can be in part parameterized by the effective number
of neutrinos $\nueff$ ({\it i.e.} the expansion rate), and in the weak
physics which determine the neutron-to-proton ratio.  In many works
$\nueff$ is regarded as the sole determinant of neutrino effects in
BBN.  However, we find that the effect of neutrino mixing on BBN is
not well described by this one parameter.  Characterizing the effects
of neutrino mixing on BBN by $\nueff$ can become overly complicated,
since the parameter has been used to describe several distinct
physical effects, including lepton number asymmetry and change in
energy density.  Another feature we find is that matter-enhanced
mixing of neutrinos in the Early Universe can alter $\nu_e$ spectra to
be non-thermal, which uniquely affects the nucleon weak rates by
lifting Fermi blocking of neutron decay by neutrinos and suppression
of neutrino capture on neutrons.

The observation of the abundances of primordial elements has greatly
increased in precision within the past few years
\cite{schburl,burtyt98a,burtyt98b,oss,bonmol}.  In SBBN, each observed
primordial abundance corresponds to a value of the baryon-to-photon
ratio $\eta \approx 2.79\times 10^8\,\Omega_b^{-1}h^{-2}$, where
$\Omega_b$ is the fractional contribution of baryon rest mass to the
closure density and $h$ is the Hubble parameter in units of $100\rm
km\,s^{-1}\,Mpc^{-1}$.  The theory is apparently successful since the
inferred primordial abundances correspond within errors to a single
value of $\eta$.  SBBN took a new turn in 1996 when it was found that
the deuterium abundance at high redshifts may be significantly lower
than that inferred through the cosmic helium abundance \cite{dtxfsb}.
The advent of high-resolution spectroscopy of quasars allowed the
measurement of deuterium abundance in high-redshift clouds between us
and distant quasars.  The ``early days'' of these high-redshift
spectral deuterium measurements were filled with controversy due to a
disparity in the inferred deuterium to hydrogen (D/H) ratio between
two observational groups working with different quasars.  More
recently, the number of quasars with low D/H has grown.  The deuterium
abundance has converged to $\sim$10\% precision: $({\rm D/H})_p \simeq (3.4
\pm 0.25) \times 10^{-5}$.  The $\eta$ value inferred from this
observation is $\eta_D \simeq (5.1 \pm 0.5) \times 10^{-10}$. The
prediction for the corresponding primordial $^4$He mass fraction is
$Y_p(D) \simeq 0.246 \pm 0.0014$.

Based on stellar absorbtion features in metal-poor HII regions, the
observed helium abundance has been argued by Olive, Steigman, and
Skillman (OSS) \cite{oss} to be $Y_p(OSS) \simeq 0.234 \pm 0.002 ({\rm
stat}) \pm 0.005 ({\rm sys})$.  The corresponding baryon-to-photon
ratio is $\eta_{OSS} \simeq (2.1 \pm 0.6) \times 10^{-10}$. The
discrepancy between the central values of $Y_p(D)$ predicted by the
Burles and Tytler deuterium observations and $Y_p(OSS)$ has led to
much speculation on non-standard BBN scenarios which could reconcile
the two measurements by decreasing the $Y_p(D)$ predicted by BBN.
However, the systematic and statistical uncertainty in the $Y_p(D)$
and $Y_p(OSS)$ puts them well within their $2\sigma$ errors.
Furthermore, Izotov and Thuan \cite{izotov} exclude a few potentially
tainted HII regions used in OSS from their analysis and arrive at a
significantly higher $Y_p \simeq 0.244 \pm 0.002 ({\rm stat})$. The
systematic uncertainty claimed in observed $Y_p$ may also be
underestimated \cite{matfulboyd}.

It has been suggested that this dubious $Y_p$ and D/H discrepancy with
standard BBN could be hinting at the presence of sterile neutrino
mixing in the Early Universe \cite{footvolk}.  However, we find that
the overall effect of active-sterile mixing in the Early Universe
would only increase $Y_p$ and thus increase the possible disparity.

Active-sterile mixing can produce an electron lepton number asymmetry
$L(\nu_e)$ before and during BBN \cite{footvolk,xdchaos}.  If the sign
of $L(\nu_e)$ is the same throughout the Universe, the helium
abundance would change to be more favorable ($\delta Y_p < 0$) or less
favorable ($\delta Y_p > 0$) with observation.  Taking into account
the effects of energy density, lepton number generation and alteration
of neutrino spectra, we find that $\delta Y_p$ can be large in the
positive direction, but limited in the negative direction \cite{lept}.
In particular, we find that matter-enhanced transformations can leave
non-thermal $\nu_e$ spectra which significantly affect the weak rates
through the alleviation of Fermi-blocking and suppression of neutrino
capture on neutrons \cite{asf}.  Furthermore, causality limits the
size of the region with a certain sign of $L(\nu_e)$ to be within the
size of the horizon during these early times ($\sim 10^{10}\;\rm cm$
at weak freeze-out) \cite{xdcaus}.  The sign of the asymmetry will
vary between regions, and the net effect of transformation on $Y_p$ will
then be the average of both the $L(\nu_e) > 0$ and $L(\nu_e) < 0$ BBN
yields, which leaves $\delta Y_p > 0$.

\section{Matter-Enhanced Mixing of Neutrinos and Neutrino Spectra\\ 
in the Early Universe}

Two possible schemes exist for producing a nonzero $L(\nu_e)$
involving matter-enhanced (MSW) transformation of neutrinos with
sterile neutrinos.  These two schemes differ greatly in the energy
distribution of the asymmetry in the spectra of the
$\nu_e/\bar{\nu}_e$ the asymmetry.  The resonant transformation of one
neutrino flavor to another depends on the energy of the neutrino.  In
the case of BBN, the active neutrinos all start off as thermal, and
sterile neutrinos are not present.  As the Universe expands, decreases
in density and cools, the position of the MSW resonance moves in
energy space from lower to higher neutrino energies.  The energy of
the resonance is \cite{lept}
\begin{equation}
\left({E\over T}\right)_{\rm res} \approx {| \delta m^2|/{\rm eV^2}
\over 16 (T/{\rm MeV})^4 L(\nu_\alpha)}
\end{equation}
where $(E/T)_{\rm res}$ is the neutrino energy normalized by the
ambient temperature $T$.  The asymmetry $L(\nu_\alpha)$ is generated as
the resonance energy moves up the neutrino spectrum.  As the Universe
cools, the resonance energy increases.  This produces a distortion of
the neutrino or anti-neutrino spectrum.  There are two plausible cases
for producing an asymmetry $L(\nu_e)$ that would affect the production
of primordial Helium through the weak rates:

\begin{itemize}

\item {\bf Direct Two Neutrino Mixing:} An asymmetry in $L(\nu_e)$ is
created directly through a $\nu_e \rightarrow \nu_s$ or $\bar{\nu}_e
\rightarrow \bar{\nu}_s$ resonance.  In this case, the resonance starts at
low temperatures for low neutrino energies, when neutrino-scattering
processes are very slow at re-thermalization.  This case leaves the
$\nu_e$ or $\bar{\nu}_e$ spectrum distorted, with the spectral
distortion from a thermal Fermi-Dirac form residing in the low energy
portion of the spectrum.  The position of the distortion cut-off moves
through the spectrum before and during BBN.  In calculating the
effects of this scenario on BBN, we included the evolving non-thermal
nature of the neutrino spectrum.

\item {\bf Indirect Three Neutrino Mixing:} An asymmetry in
$L(\nu_\tau)$ or $L(\nu_\mu)$ created by a $\nu_\tau \rightarrow \nu_s
(\nu_\mu \rightarrow \nu_s)$ resonance is later transferred into the
$L(\nu_e)$ through a $\nu_\tau \rightarrow \nu_e (\nu_\mu \rightarrow
\nu_e)$ resonance ($L(\nu_e) > 0$). This will happen for the
anti-neutrino flavors for the $L(\nu_e) < 0$ case.  In this scenario,
the resonant transitions occur at higher temperatures when
neutrino-scattering is more efficient.  Both the $\nu_\tau (\nu_\mu)$
and $\nu_e$ spectra re-thermalize, but since the temperature is below
weak decoupling, the asymmetry in neutrino lepton numbers remain.  The
effect of this scenario on BBN can be described by suppression or
enhancement over the entire $\nu_e/\bar{\nu}_e$ spectra.

\end{itemize}

\section{Neutron-Proton Interconversion Rates and Neutrino Spectra}

BBN can be approximated as the freeze-out of nuclides from nuclear
statistical equilibrium in an ``expanding box.''  Nuclear reactions
freeze-out, or fail to convert one nuclide to another when their rate
falls below the expansion rate of the box (the Hubble expansion).  As
different reactions rates freeze-out at different temperatures, the
abundances of various nuclides get altered and then become fixed.  The
nuclide produced in greatest abundance (other than hydrogen) is
$^4$He.  The reactions that affect the $^4$He abundance ($Y_p$) the
most are the weak nucleon interconversion reactions
\begin{equation}
\label{reactions}
n + \nu_e \leftrightarrow p + e^- \;\;\;\;\; 
n + e^+ \leftrightarrow p + \bar{\nu}_e \;\;\;\;\;
n \leftrightarrow p + e^- + \bar{\nu}_e .
\end{equation}
These reactions all depend heavily on the density of the electron-type
neutrinos and their energy distributions \cite{alpher}.  The two
neutrino transformation scenarios for altering the $\nu_e$ spectrum
described above affect $Y_p$ through these rates and through the
effect of increased energy density on the expansion rate.

\subsection{Direct Two-Neutrino Mixing: $\nu_s \leftrightarrow \nu_e$}

In this situation, either the populations of the $\nu_e$ or
$\bar{\nu}_e$ energy spectra are cutoff at lower energies.  The
reaction rates that alter the overall $n\leftrightarrow p$ rates the
most because of the spectral cutoff are
\begin{equation}
\label{rate1}
\lambda_{n\rightarrow p e \nu} = A \int{v_e E_\nu^2 E_e^2 dp_\nu
[1+e^{-E_\nu/k T_\nu}]^{-1} [1+e^{-E_e/k T_\nu}]^{-1}}
\end {equation}
\begin{equation}
\label{rate2}
\lambda_{n \nu \rightarrow p e} = A \int{v_e E_e^2 p_\nu^2 dp_\nu
[e^{E_\nu/k T_\nu} + 1]^{-1} [1+e^{-E_e/k T_\nu}]^{-1}}
\end {equation}
(the notation in these expressions follows that in
Ref. \cite{weinberg}). The rate in Eq. \ref{rate1} is modified with
the $\bar{\nu}_e$ spectra, and the rate in Eq. \ref{rate2} is modified
with the $\nu_e$ spectra.

\begin{figure}[ht]      
\centerline{\epsfxsize 5 truein \epsfbox{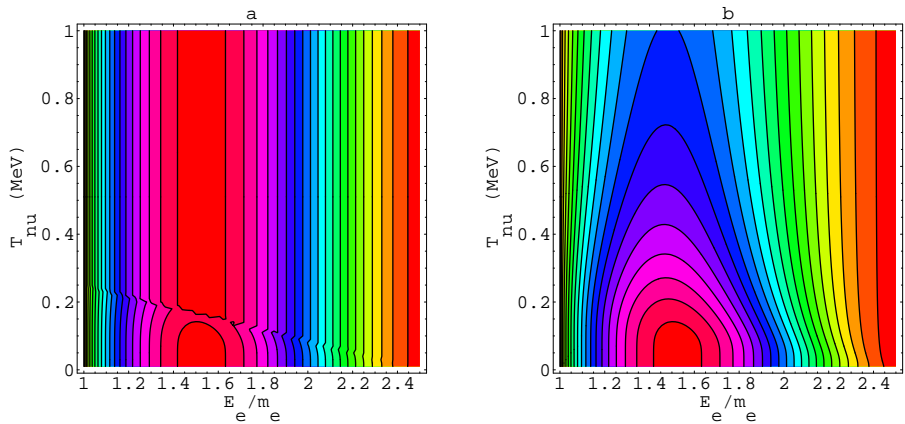}}   
\vskip 0 cm
\caption[]{
\label{intnueb}
\small 
Plotted are the contours of the amplitude of the rate
integrand for $n \rightarrow p + e^- + \bar{\nu}_e$, with increasing
amplitude towards the center. (a) shows the lack of fermi blocking by
low energy electron anti-neutrinos at higher temperatures.  (b) shows
the standard BBN integrand.}
\end{figure}

\begin{figure}[ht]      
\centerline{\epsfxsize 5 truein \epsfbox{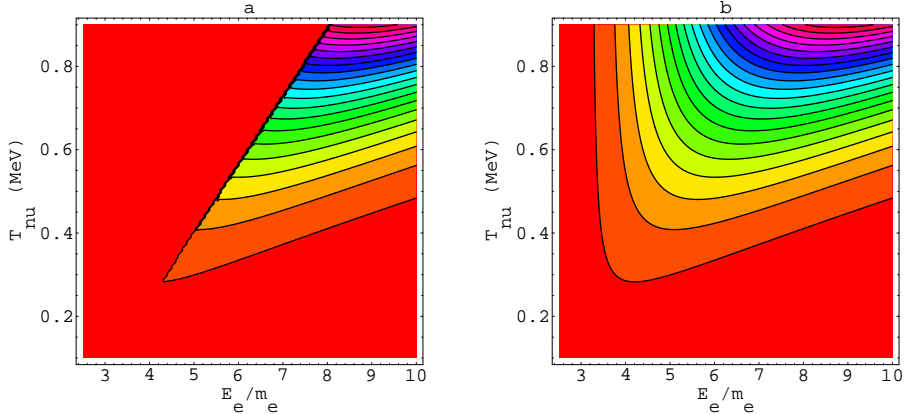}}   
\vskip 0 cm
\caption[]{
\label{intnue}
\small Plotted are the contours of the amplitude of the rate
integrand for $n + \nu_e \rightarrow p + e^-$, with increasing amplitude
towards the upper right: (a) shows the blocking of the integrand for lower
energies due to the suppression of low energy electron neutrinos;  (b) shows
the standard BBN integrand.}
\end{figure}

For a spectral cutoff for $\bar{\nu}_e$, corresponding to
$\bar{\nu}_{s}\leftrightarrow \bar{\nu}_e$ transformations, the
Fermi-blocking term $[1+e^{-E_\nu/k T_\nu}]^{-1}$ in neutron decay
(Eq. \ref{rate1}) becomes unity at low $\bar{\nu}_e$ energies.  The
rate integrand is then greatly enhanced when low energy neutrinos
would otherwise block the process (see Figure \ref{intnueb}).  In the
case of a spectral cutoff for $\nu_e$, corresponding to
$\nu_{s}\leftrightarrow \nu_e$ transformations, the spectral term
$p^2_\nu[e^{E_\nu/k T_\nu} + 1]^{-1}$ and thus the rate integrand go
to zero for low energy neutrinos (see Fig \ref{intnue}).

\begin{figure}[ht]      
\centerline{\epsfxsize 3.5 truein \epsfbox{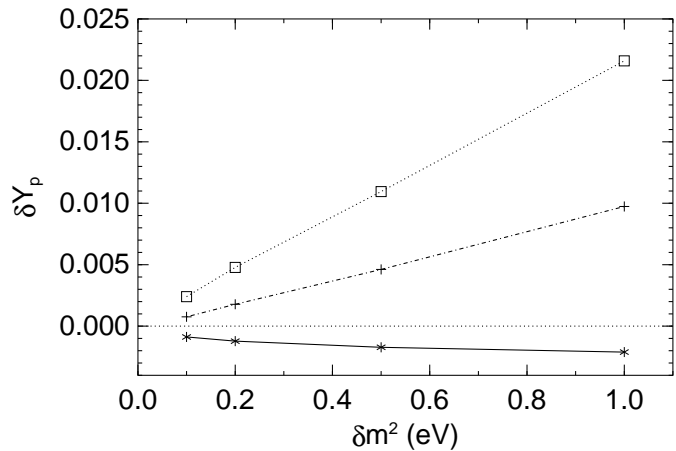}}   
\vskip 0 cm
\caption[]{
\label{dycut}
\small 
Plotted are $\delta Y_p$ vs. $\delta m^2$ for the direct two-neutrino
transformation scenario.  The top line is for $L(\nu_e) < 0$; the bottom
is for $L(\nu_e) > 0$; the middle line is the average of both cases
enforced by causality.
}
\end{figure}

The resultant change in $Y_p$ for this ``two neutrino'' case is shown
in Figure \ref{dycut}.  For $\bar{\nu}_e \rightarrow \bar{\nu}_s$
transformations, $L(\nu_e) > 0$, the neutron decay rate in
Eq. \ref{rate1} is enhanced, fewer neutrons are present, and $Y_p$ is
decreased.  For $\nu_e \rightarrow \nu_s$ transformations, $L(\nu_e) <
0$, the $n\,\nu_e\rightarrow p\, e$ rate in Eq. \ref{rate2} is
suppressed, more neutrons are present, and $Y_p$ is increased.  The
effect of the rate in Eq. \ref{rate2} is much greater than that of the
rate in Eq. \ref{rate1} on the neutron-to-proton ratio and $Y_p$. This
causes a larger magnitude change in the neutron-to-proton ratio for the
$L(\nu_e) < 0$ case, and thus a larger magnitude change in $Y_p$ in
the positive direction than in the negative.

The sign of the asymmetry $L(\nu_e)$ is randomly determined
\cite{xdchaos}.  Because of causality, different horizons will have an
asymmetry of random sign \cite{xdcaus}, fifty-percent positive and
fifty-percent negative.  The effect of the asymmetry on $Y_p$ will
then be the average of these two cases, which is also plotted in
Figure \ref{dycut}.

\subsection{Indirect Three-Neutrino Mixing}

\begin{figure}[ht]      
\centerline{\epsfxsize 3.5 truein \epsfbox{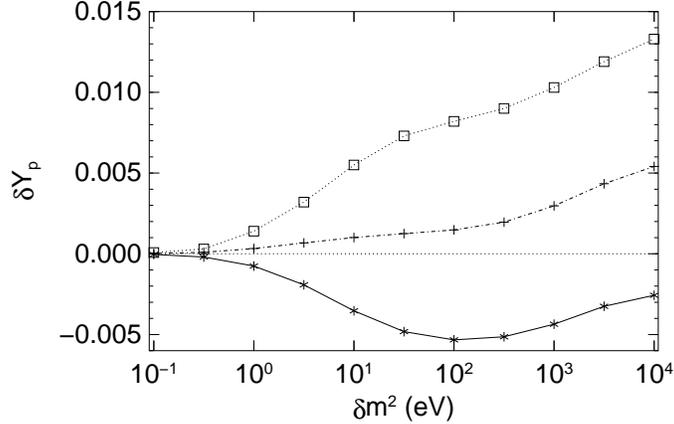}}   
\vskip 0 cm
\caption[]{
\label{dyoverall}
\small 
Plotted are $\delta Y_p$ vs. $\delta m^2$ for the indirect three-neutrino
transformation scenario.  The top line is for $L(\nu_e) < 0$; the bottom
is for $L(\nu_e) > 0$; the middle line is the average of both cases
enforced by causality.
}
\end{figure}

In the case that an asymmetry is produced in $\nu_e/\bar{\nu}_e$
indirectly through an initial transformation between $\nu_\tau
\rightarrow \nu_s$ or $\nu_\mu \rightarrow \nu_s$, the neutrinos can
re-thermalize through self-scattering at higher temperatures.  The
number densities remain asymmetric, but the asymmetry is spread
uniformly over the entire energy spectra. The effect on the weak rates
is not as drastic as the above case, but is still present. In this
case, the increase in energy density due to the thermalization of
sterile neutrinos also affects $Y_p$ through the increase in the
expansion rate.  Thus, for $L(\nu_e) > 0$, the suppression of $Y_p$ is
counteracted by the increased expansion rate.  The increase in energy
density always causes a positive change in $Y_p$, while the change
caused by the weak nucleon rates depends on the sign of $L(\nu_e)$,
thus the overall resulting change in $Y_p$ is greater in the positive
direction than in the negative (see Figure \ref{dyoverall}). For the
$L(\nu_e) > 0$ case, the negative change $\delta Y_p$ has a minimum at
$\delta m^2 \approx 100\,\rm eV^2$ since at this point the positive
effect of energy density begins to dominate the negative effect of the
modification of the weak rates.

\section{Conclusions}

The characterization of the effects of neutrino transformations on BBN
through a single parameter $N_\nu$ hides varied physical effects of
neutrino mass and mixing that are non-trivially related.  These
effects include an increase in energy density and modification in the
nucleon weak rates.  The modification of these rates is through either
an asymmetry that is distributed over the entire spectrum, or an
asymmetry confined to low energy $\nu_e/\bar{\nu}_e$.  We find that
the distribution of the $L(\nu_e)$ in the spectra is important to the
weak rates, the neutron-to-proton ratio, and the production of $^4$He.
In addition, the change in the predicted $Y_p$ is always greater in
the positive direction than in the negative, thus when averaged over
causal horizons, $\delta Y_p$ is always positive. The changes $\delta
Y_p$ are comparable to the uncertainty in current $Y_p$ measurements,
thus resonant neutrino mixing can measurably affect the BBN
predictions.  It is important to note that there may remain
non-trivial portions of parameter space in non-standard
BBN---including, but not limited to, neutrino mixing, spatial
variations in $\eta$, and massive decaying particles---that fit within
the observed uncertainties of the primordial abundances.

K.~A., X.~S. and G.~M.~F. are partially supported by NSF grant
PHY98-00980 at UCSD.

\end{document}